\documentstyle{article}
\begin{document}
%\baselineskip 0.33in
%\sloppy
%\begin{center}
\centerline{\large {\bf Particle-hopping Models of Vehicular Traffic:}} 
\centerline{\large {\bf Distributions of Distance Headways and}}
\centerline{\large {\bf Distance Between Jams}}
\vspace{2cm}
\centerline{\bf Debashish Chowdhury$^{1,2}$, Kingshuk Ghosh$^1$, Arnab Majumdar$^1$,}
\centerline{\bf Shishir Sinha$^1$ and R. B. Stinchcombe$^3$} 
\vspace*{1cm}
\centerline{$^1$Physics Department, Indian Institute of Technology} 
\centerline{Kanpur 208016, India}
\vspace*{.5cm}
\centerline{$^2$Institute for Theoretical Physics, University of Cologne,}
\centerline{D-50937 K\"oln, Germany}
\vspace*{.5cm}
\centerline{$^3$Department of Theoretical Physics, University of Oxford} 
\centerline{Oxford Ox1 3NP, England}
%\end{center} 
\begin{abstract}
We calculate the distribution of the distance headways (i.e., the 
instantaneous gap between successive vehicles) as well as the distribution 
of instantaneous distance between successive jams in the Nagel-Schreckenberg 
(NS) model of vehicular traffic. When the maximum allowed speed, $V_{max}$, 
of the vehicles is larger than unity, over an intermediate range of 
densities of vehicles, our Monte Carlo (MC) data for the distance headway 
distribution exhibit two peaks, which indicate the coexistence of 
"free-flowing" traffic and traffic jams. Our analytical arguments clearly 
rule out the possibility of occurrence of more than one peak in the 
distribution of distance headways in the NS model when $V_{max} = 1$  as  
well as in the asymmetric simple exclusion process. Modifying and extending 
an earlier analytical approach for the NS model with $V_{max} = 1$, and 
introducing a novel transfer matrix technique, we also calculate the exact 
analytical expression for the distribution of distance between the jams in 
this model; the corresponding distributions for $V_{max} > 1$ have been 
computed numerically through MC simulation.  

\end{abstract}
\vspace{0.5in}



\vfill\eject 

{\bf 1. INTRODUCTION:}

The motion of each vehicle in traffic is influenced by others around it; 
the statistics and dynamics of traffic flow [1-5] crucially depend on  
these "interactions" between the vehicles. However, the dynamics of vehicles is 
more complicated than that of interacting particles [6] because of the 
dependence on human behaviour; usually, the driver has the $aim$ of 
reaching his destination safely in the shortest time and can also $learn$ 
from experience. Nevertheless, significant progress has been made in 
modelling traffic by extending the concepts and techniques of statistical 
mechanics [4,5] by capturing some of the common and essential features of 
the driving habits of the individual drivers. 

The theoretical models of traffic can be broadly divided into two classes: 
(i) $continuum$ models, and (ii) $car-following$ models; the former 
is the analogue of the "hydrodynamic" models of fluids while the latter is the 
analogue of the "microscopic" models in statistical mechanics. The older 
version of the car-following theory [7] was formulated mathematically 
using differential equations for describing the time evolution of the 
positions and speeds of the individual vehicles in time, which was treated 
as a continuous real variable. In the modern versions, namely, the $particle-
hopping$ models, not only the position and speed of the vehicles but time 
also is assumed to be discrete. Some of these particle-hopping models [8] 
may be regarded as stochastic cellular automata [9]. 

The $distance$ $headway$ is one of the most important characteristics of 
vehicular traffic. It is defined as the distance from a selected point on 
the lead vehicle to the same point on the following vehicle. Usually, the 
front edges or bumpers are selected [3]. One of the aims of this paper  
is to calculate the distributions of the distance headways in the steady state, 
using a class of particle-hopping models, which are known to capture some 
of the essential features of traffic flow on highways. We also report 
analytical as well as numerical results on the distribution of instantaneous  
distance between successive vehicular jams in the steady state of these models. 

We define the models and the characteristic quantities of interest in 
section 2. We compute the distance headway distribution in section 3 
and the distribution of distance between jams in section 4. We summarize 
the results and conclusion in section 5. 

{\bf 2. THE MODELS AND CHARACTERISTIC QUANTITIES OF INTEREST:}

In all the "microscopic" models under our consideration, a lane is 
represented by a one-dimensional lattice of $L$ sites (i.e., a chain of 
$L$ equispaced points). Each of the lattice sites can be either empty or 
occupied by a "vehicle". There is an "exclusion principle", namely, that 
no two vehicles can occupy the same lattice site simultaneously.  

The ratio of the number of occupied lattice sites to the total number of 
lattice sites is defined as the density of the vehicles on the lane. The 
density is time-dependent for a finite lattice with open boundaries where 
vehicles enter from one end and leave from the other. In this paper  
we shall consider only idealized single-lane highways with 
periodic boundary conditions.  Therefore, in this geometry 
of the highway, the time-independent density $c$ of the vehicles is $N/L$ where 
$N (\leq L)$ is the total number of vehicles in the traffic . 

{\bf 2.1. The Asymmetric Simple Exclusion Process:} 

One of the simplest dynamical models of interacting particle systems is 
the so-called {\it asymmetric simple exclusion process} (ASEP) which may 
be regarded as a caricature of vehicular traffic.  
The rules for updating the positions of the particles in the ASEP are 
as follows: one particle is picked at random and moved forward by 
one lattice site if the new site is empty. In this random sequential 
update, it is the random picking that introduces stochasticity (noise) 
into the model. Although speeds of the particles do not explicitly enter 
into the rules, the effective speed of a particle in the ASEP can take 
only two values, namely, $0$ and $V_{max} = 1$. 

{\bf 2.2. The Nagel-Schreckenberg Model:} 

We now recall the "rules" describing the dynamics of vehicles in the 
Nagel-Schreckenberg (NS) model [8] where the speed $V$ of each vehicle 
can take one of the $V_{max}+1$ allowed integer values  $V=0,1,...,V_{max}$. 
At each discrete time step $t \rightarrow t+1$, the arrangement of $N$ 
vehicles is updated {\it in parallel} according to the following rules.
Suppose, $V_n$ is the speed of the $n$-th vehicle at time $t$. 

{\it Step 1: Acceleration.} If, $ V_n < V_{max}$, the speed 
of the $n$-th vehicle is increased by one, i.e., $V_n \rightarrow V_n+1$.

{\it Step 2: Deceleration (due to other vehicles).} If $d$ is the gap
in between the $n$-th vehicle and the vehicle in front of it, and if 
$d \le V_n$, the speed of the $n$-th vehicle is reduced to $d-1$, i.e., 
$V_n \rightarrow d-1$.

{\it Step 3: Randomization.} If $V_n > 0$, the speed of the $n$-th vehicle 
is decreased randomly by unity (i.e., $V_n \rightarrow V_n-1$) with 
probability $p$ ($0 \leq p \leq 1$); $p$, the random deceleration probability, 
is identical for all the vehicles and does not change during the updating.

{\it Step 4: Vehicle movement.} Each vehicle is moved forward so that 
$X_n \rightarrow  X_n + V_n$ where $X_n$ denotes the position of the 
$n$-th vehicle at time $t$.  

In the literature on traffic flow, the relation between the average flux 
of the vehicles and their density is usually referred to as the "fundamental  
relation". The qualitative features of the fundamental relation of the 
NS model are very similar to those of the empirically derived fundamental 
relation for real traffic. $c_m$, the density corresponding to the 
maximum flux, is usually called the $optimum$ density. 

{\bf 2.3. Definitions of Distance Headway and Distance Between Jams:} 

The number of empty lattice sites in front of a vehicle is taken to be
the measure of the corresponding distance headway. If the 
instantaneous speed of a vehicle is zero, then it is part of a traffic 
jam. The number of lattice sites in between two such "jammed" vehicles 
is taken as the measure of the distance between the two corresponding 
instantaneous jams to which they belong, provided none of the sites 
between these two sites is occupied by any other "jammed" vehicle (the 
sites in between may, of course, be either empty or occupied by vehicles 
with non-zero instantaneous speed).  

We have calculated both the distributions for the NS model analytically 
when $V_{max} =1$. However, for the NS model with $V_{max} > 1$ our 
results in this paper have been obtained by computer simulation as 
analytical calculations become too complicated to carry through. 

{\bf 3. RESULTS ON DISTANCE HEADWAY DISTRIBUTION:} 

{\bf 3.1. Distance Headway Distribution for $V_{max} > 1$:} 

NS claimed that $V_{max}=5$ is a very realistic choice and, therefore, 
most of the subsequent works on the NS model have been carried out for 
this particular value of $V_{max}$. Besides, the most common choice 
for $p$ is $0.5$. Therefore, we begin presentation 
of our results by plotting the distribution of distance headways in 
the NS model in fig.1(a) for $V_{max} = 5$, $p = 0.5$. 
For small values of the density $c$, the gap distribution exhibits a 
single peak; interestingly, the gap distribution vanishes for gap sizes 
smaller than the corresponding value of $V_{max}$. This implies that, at 
low densities, the vehicles arrange themselves in such a manner as to 
maintain a gap, which is at least as large as $V_{max}$, 
so that free flow of the vehicles can take place. With the increase 
of the density of the vehicles the distribution still exhibits a single 
maximum, although the most probable gap becomes smaller, provided the 
density is sufficiently low. However, as $c$ approaches $c_m$, a second peak 
appears at a gap of unit lattice spacing; this new peak is a consequence of 
the congestion of the vehicles. With the further increase of density the new 
peak becomes higher at the cost of that at the larger gap and, eventually, 
as $c \rightarrow 1$, the distribution approaches a $\delta$-function at unity. 
 
Are the characteristics of the distribution in fig.1(a) merely special 
feature of the chosen value $V_{max} = 5$ or generic features of the NS 
model irrespective of the specific value of $V_{max}$? In order to answer 
this question let us  plot the distribution of distance headways in the 
NS model for another value of $V_{max}$, say, $V_{max} = 3$ in fig.1(b) 
where $p$ is chosen to be $0.5$ as in fig.1(a). The characteristics of 
the distribution and the trend of its variation with density are very 
similar to those in fig.1(a). 

Now we shall show that the occurrence of two peaks in the distribution 
of distance headways depends on $p$. For example, for the same $p$ as 
in fig.1, namely, $p = 0.5$, the distribution does not show the 
occurrence of two peaks for any value of the density $c$ when $V_{max}$ 
is $2$ (see fig.2(a)). Nevertheless, the two-peak structure of the 
distribution over a regime of density is observed even for $V_{max} = 2$ 
provided $p$ is large enough (see fig.2(b)).

{\bf 3.2. Distance Headway Distribution for $V_{max} = 1$:} 
 
Does the distance headway distribution in the NS model exhibit two peaks 
also for $V_{max} = 1$? In order to establish the absence of any such 
two-peak structure in the gap distributions, irrespective of the density 
of the vehicles, in the NS model for $V_{max} = 1$ and in the ASEP we 
next derive the corresponding exact analytical expressions. 

For the convenience of their analytical calculations, Schreckenberg 
et al.[10] changed the order of the steps in the update rules in such a 
manner that it does not influence the steady-state properties of the model. 
They assumed the sequence of steps $2-3-4-1$, instead of $1-2-3-4$; the 
advantage is that there is no vehicle with $V = 0$ immediately after 
the acceleration step and this reduces the number of possible states of 
a site by one. For example, if $V_{max} = 1$, $V$ can take only the 
value $1$ at the end of a sequence of steps $2-3-4-1$ and Schreckenberg 
et al.[10] could use a binary site variable $\sigma$ to describe the 
state of each site; $\sigma = 0$ represented an empty site and 
$\sigma = 1$ represented a site occupied by a vehicle (with speed $V=1$). 
In this notation, the probability of finding an $n-cluster$ in the 
configuration $(\sigma_1, \cdots ,\sigma_n)$ in the steady state can 
be denoted by the symbol $P_n(\sigma_1, \cdots ,\sigma_n)$. 

Although the sequence of steps $2-3-4-1$ and the description of the 
configurations in terms of the binary variable $\sigma$ can be used 
also for our analytical calculation of the distance headway distribution 
in the NS model with $V_{max} = 1$ these, however, would be indequate for 
our analytical calculation of the distribution of distance between jams 
in the same model, as we shall explain in the next section. Therefore, 
for the convenience of extension of the approach in the next section 
we now slightly modify the notation. We use the site variable $s$, instead 
of $\sigma$, to denote the state of each site; $s = -1$ represents an 
empty site whereas $s = 1$ represents a site occupied by a vehicle 
(with speed $V = 1$). But, for the calculation of the distance headway 
distribution, we adopt the same sequence as in ref.[10], i.e., $2-3-4-1$. 
Therefore, in this section, the site variable $s$ can take either of 
the only two allowed values, namely, $+1$ and $-1$.   

Suppose, ${\cal P}(j)$ is the probability of finding a gap of size $j$ in 
front of a vehicle. Stated more precisely, ${\cal P}(j)$ is the conditional 
probability of finding $j$ empty sites in front of a site which is given 
to be occupied by a vehicle, i.e., 
\begin{equation}
{\cal P}(j) = P(\underline{1}|\underbrace{-1 -1 \cdots -1}_{j}1) 
\end{equation} 
where $P$ is the probability of occurrence of the argument configuration 
where the underlined part is fixed. 

In the mean field approximation we neglect all correlations between the 
vehicles. Therefore, in this approximation, the probability on the 
right hand side of equation (1) can be factorized and, hence, the gap 
distribution is
\begin{equation}
{\cal P}_{mfa}(j) = c(1-c)^{j} \; \; \mbox{ for $j = 0,1,2,...$}
\end{equation}
Mean-field approximation is known to be exact for the ASEP [11-13] 
and, therefore, the right hand side of equation (1)  is the exact 
analytical expression for gap distribution between the particles in 
the ASEP; note that ${\cal P}_{mfa}(j)$ does not exhibit two peaks 
simultaneously at any value of $c$ (see fig.3). 

Mean field results can be improved by taking into account the correlation 
between the vehicles. In general, this is achieved by dividing the lattice 
into clusters of length $n$ which amounts to taking the effect of correlation 
upto $n$ nearest neighbours [10]. In the $n$-cluster approximation, we divide 
the lattice into segments or clusters of length $n$, such that two 
neighbouring clusters have $(n-1)$ sites in common. For the case of 
$V_{max} = 1$ one expects strong correlation only between vehicles at 
nearest-neighbout lattice sites and, therefore, from now onwards, we carry 
out all our analytical calculations for this case only in the 2-cluster 
approximation. 

In the 2-cluster approximation [10], the probability of finding a gap 
of $j$ sites, can now be written as [14]
\begin{equation}
{\cal P}_{2c}(j) = P_2(\underline{+1}|+1) \quad for \quad j = 0, 
\end{equation}
\begin{equation}
{\cal P}_{2c}(j) = P_2(\underline{+1}|-1) P_2(\underline{-1}|+1), \quad for \quad j = 1, 
\end{equation}
\begin{equation}
{\cal P}_{2c}(j) = P_2(\underline{+1}|-1) \{P_2(\underline{-1}|-1)\}^{j-1} P_2(\underline{-1}|+1), \quad for \quad j \geq 1, 
\end{equation}
where the conditional probabilities $P_2(s_{i-1}|\underline{s_i})$ and 
$P_2(\underline{s_{i+1}}|s_{i+2})$ are given, interms of the 
2-cluster probabilities $P_2(s_{i-1},s_i)$ and $P_2(s_{i+1},s_{i+2})$, 
by the relations 
\begin{equation}
P_2(s_{i-1}|\underline{s_i}) = \frac{P_2(s_{i-1},s_i)}{P_2(-1,s_i) + P_2(1,s_i)} 
\end{equation} 
and   
\begin{equation}
P_2(\underline{s_{i+1}}|s_{i+2}) = \frac{P_2(s_{i+1},s_{i+2})}{P_2(s_{i+1},-1) + P_2(s_{i+1},1)} 
\end{equation} 

Schreckenberg et al.[10] pointed out that only one of the four   
2-cluster probabilities $P_(+1,+1), P_2(+1,-1), P_2(-1,+1), P_2(-1,-1)$ 
is independent. Solving the master equation for $P_2(+1,-1)$ they 
found that   
\begin{equation}
P_2(+1,-1) = \left(\frac{1}{2q}\right)[1 - \{1 - 4qc(1-c)\}^{1/2}] = y
\end{equation}
where $q = 1 - p$. The three other 2-cluster probabilities can be 
obtained from the equations 
\begin{equation}
P_2(+1,+1) = c - y 
\end{equation}
\begin{equation}
P_2(-1,+1) = y 
\end{equation}
\begin{equation}
P_2(-1,-1) = 1 - c - y 
\end{equation}
Using the expressions (6)-(11) in the equations (3)-(5) we get [14] 
\begin{equation}
\begin{array}{rcll}
{\cal P}_{2c} (j) &=& 1 - (y/c) &  \mbox{ for $j = 0$}\\
                  &=& \{{y^{2} \over c(1-c)}\}\left(1 - {y \over (1-c)}
                  \right)^{j-1} &\mbox{ for $j = 1,2,3,...$}
\end{array}
\end{equation}
where $y = P_2(+1,-1)$. 
The distribution (12) is the exact analytical expression for the distance 
headway distribution in the NS model when $V_{max} = 1$ and it also 
does not exhibit two peaks simultaneously at any value of $c$ (see fig.4).  
The differences between the exact gap distributions for the ASEP and the 
NS model with $V_{max} = 1$ arise from the fact that parallel updating in 
used in the NS model in contrast to the random sequential updating in the 
ASEP [8,10]. 

The two-peak structure of the gap distribution has been interpreted as 
a signature of "two-phase coexistence" in a recent study of a continuum model 
[15]. We believe that an identical interpretation can be given also for the 
occurrence of two-peak structure in the gap distribution of the discrete NS 
model; the two coexisting phases being the "free-flowing phase" and a "jammed" 
phase. In the context of particle-hopping models, the two-peak structure of 
the gap distribution was first discovered during a computer simulation of 
a two-lane model [16] but no analytical calculations were attempted. Our 
comparison here with the exact analytical expressions for $V_{max} = 1$ 
clearly establishes qualitative differences between the distributions for 
$V_{max} = 1$ and those for $V_{max} > 1$; there is only one peak 
in the gap distribution for all densities when $V_{max} = 1$ 
whereas for $V_{max} > 1$ two peaks appear in an intermediate regime of 
density of vehicles. In this paper we have indirectly inferred the 
coexistence of the two dynamical phases in the NS model with $V_{max} > 1$ 
over a $p$-dependent range of densities by computing the distance 
headway distribution which is, traditionally, one of the important 
characteristics of vehicular traffic.  Very recently, L\"ubeck et al.[17] 
have proposed a new method, based on the "local density distribution",  
in an attempt to characterize the free-flowing and jammed phases 
quantitatively and to locate the region of coexistence of these phases on 
a "phase diagram" of the NS model.

{\bf 4. RESULTS ON THE DISTRIBUTION OF DISTANCE BETWEEN JAMS:} 

{\bf 4.1. Distribution of Distances between Jams for $V_{max} = 1$:} 

According to our definition, each of the jammed vehicles has a speed 
$V = 0$. Therefore, for the calculation of the distribution of distance 
between the jams the site variable $s$ must be allowed to take one of the 
{\it three} allowed values, namely, $s = -1$ if the site is empty, $s = 1$ 
if the site is occupied a vehicle whose instantaneous speed is $V = 1$ and 
$s = 0$ if the site is occupied by a vehicle which is instantaneously 
jammed, i.e., has a speed is $V = 0$. Since $s$ can now take three 
values, namely, $1, 0, -1$, we have a total of nine 2-cluster probabilities. 
Therefore, in the case of the sequence $2-3-4-1$ of the steps in the 
NS model, we have the equations 
\begin{equation}
P_2(0,1) = P_2(0,0) = P_2(0, -1) = P_2(1,0) = P_2(-1,0) = 0. 
\end{equation}
in addition to the equations (8)-(11).

Since there is no vehicle with speed $V = 0$ after the step (1) of the 
NS model, we cannot use the sequence $2-3-4-1$ for the calculation of the 
distribution of distance between jams. Instead, {\it we  use 
the sequence} $4-1-2-3$, i.e., we calculate the required two cluster 
probabilities after the {\it step 3} of the NS model [18]. In order avoid any 
confusion between the two-cluster probabilities $P_2(s_i,s_{i+1})$ 
associated with the sequence $2-3-4-1$, which we considered in section 3, 
and those associated with the sequence $4-1-2-3$ we denote the latter by 
the symbol $p_2(s_i,s_{i+1})$; we shall use the expressions for 
$P_2(s,s')$ to calculate those for $p_2(s,s')$. 

In the 2-cluster approximation, the probability of a distance $k$ between 
two successive jams is given by 
\begin{equation}
\tilde{P}_{2c}(k) = \sum_{s_i=\pm1} p_2(\underline{0}|s_1) \underbrace{p_2(\underline{s_1}|s_2) \cdots}_{(k-1) times} p_2(\underline{s_k}|0) \quad for \quad k \geq 1 
\end{equation}
and 
\begin{equation}
\tilde{P}_{2c}(k) = p_2(\underline{0}|0) \quad for \quad k = 0. 
\end{equation}

What is the physical meaning of the probability of gap $k = 0$ between 
two jams? Shouldn't two jammed vehicles with vanishing gap be regarded 
part of the same jam? Indeed, the equation (15) clearly shows that 
$\tilde{P}_{2c} (k=0)$ is the total fraction of the vehicles which are 
simultaneously in the jammed state. 

The required 2-cluster probabilities $p_2(a,b)$ can be calculated from the 
expressions for $P_2(c,d)$, as expressed by the following equations [18] 
\begin{equation}
p_2(-1,-1) = P_2(-1,-1) 
\end{equation}
\begin{equation}
p_2(-1,0) = p P_3(-1,+1|-1) + P_3(-1,+1|+1) 
\end{equation}
\begin{equation}
p_2(-1,+1) = q P_3(-1,+1|-1)  
\end{equation}
\begin{equation}
p_2(0,-1) = p P_2(+1,-1)  
\end{equation}
\begin{equation}
p_2(0,0) = p P_3(+1,+1|-1) + P_3(+1,+1|+1) 
\end{equation}
\begin{equation}
p_2(0,+1) = q P_3(+1,+1|-1)  
\end{equation}
\begin{equation}
p_2(+1,-1) = q P_2(+1,-1)  
\end{equation}
\begin{equation}
p_2(+1,0) = 0   
\end{equation}
\begin{equation}
p_2(+1,+1) = 0   
\end{equation}

The new conditional probabilities defined by 
\begin{equation}
p_2(s_{i-1}|\underline{s_i}) = \frac{p_2(s_{i-1},s_i)}{p_2(-1,s_i) + p_2(0,s_i) + p_2(1,s_i)} 
\end{equation} 
and   
\begin{equation}
p_2(\underline{s_{i+1}}|s_{i+2}) = \frac{p_2(s_{i+1},s_{i+2})}{p_2(s_{i+1},-1) + p_2(s_{i+1},0) + p_2(s_{i+1},1)} 
\end{equation} 
can be obtained from the equations (16)-(24) using 2-cluster approximations 
for the 3-cluster probabilities, 
i.e., 
\begin{equation}
P_3(s,s'|s'') = P_2(s,s') P_2(\underline{s'}|s'') = P_2(s,s')\biggl( \frac{P_2(s',s'')}{P_2(s',-1)+P_2(s',+1)} \biggr) 
\end{equation}
and the expressions (8)-(11), (13), for the 2-cluster probabilities 
$P_2(s,s')$. Thus, we get 
\begin{equation}
p_2(\underline{0}|-1) = \frac{py}{c-qy} 
\end{equation}
\begin{equation}
p_2(\underline{0}|0) = 1 - \frac{y}{c} 
\end{equation}
\begin{equation}
p_2(\underline{0}|+1) = \frac{qy(c-y)}{c(c-qy)} 
\end{equation}
\begin{equation}
p_2(\underline{-1}|-1) = \frac{1-c-y}{1-c} 
\end{equation}
\begin{equation}
p_2(\underline{-1}|0) = \frac{y(c-qy)}{c(1-c)} 
\end{equation}
\begin{equation}
p_2(\underline{-1}|+1) = \frac{qy^2}{c(1-c)} 
\end{equation}
\begin{equation}
p_2(\underline{+1}|-1) = 1  
\end{equation}
\begin{equation}
p_2(\underline{+1}|0) = 0  
\end{equation}
\begin{equation}
p_2(\underline{+1}|+1) = 0  
\end{equation}
We shall use these expressions for the conditional probabilities 
in equations (14) and (15) to derive the distribution of the 
distance between the jams.

We now define the $2 \times 2$ transfer matrix $T$ whose elements 
$T[\alpha, \beta]$ ($\alpha, \beta = 1,2$) are given by [18]
\begin{equation}
T[1,1] = p_2(\underline{1}|1) = 0 
\end{equation}
\begin{equation}
T[1,2] = p_2(\underline{1}|-1) = 1  
\end{equation}
\begin{equation}
T[2,1] = p_2(\underline{-1}|1) = \frac{qy^2}{c(1-c)}   
\end{equation}
\begin{equation}
T[2,2] = p_2(\underline{-1}|-1) = \frac{1-c-y}{1-c}   
\end{equation}
Hence, within the 2-cluster approximation, 
\begin{eqnarray}
\tilde{P}_{2c}(k) = p_2(\underline{0}|1) T^{k-1}[1,1] p_2(\underline{1}|0) + 
         p_2(\underline{0}|-1) T^{k-1}[2,1] p_2(\underline{1}|0) \nonumber \\ 
         + p_2(\underline{0}|1) T^{k-1}[1,2] p_2(\underline{-1}|0) + 
         p_2(\underline{0}|-1) T^{k-1}[2,2] p_2(\underline{-1}|0)  
\end{eqnarray}
where $T^{k-1}[\alpha,\beta]$ refers to the $[\alpha,\beta]$ element 
of the matrix $T^{k-1}$. Now 
\begin{equation}
T^{k-1}[1,1] = \frac{\lambda_1 \lambda_2^{k-1} - \lambda_2 \lambda_1^{k-1}}{\lambda_1 - \lambda_2} 
\end{equation}
\begin{equation}
T^{k-1}[1,2] = \frac{\lambda_1^{k-1} - \lambda_2^{k-1}}{\lambda_1 - \lambda_2} 
\end{equation}
\begin{equation}
T^{k-1}[2,1] = \frac{\lambda_1 \lambda_2^{k} - \lambda_2 \lambda_1^{k}}{\lambda_1 - \lambda_2} 
\end{equation}
\begin{equation}
T^{k-1}[2,2] = \frac{\lambda_1^{k} - \lambda_2^{k}}{\lambda_1 - \lambda_2} 
\end{equation}
where 
\begin{equation}
\lambda_1 = \biggl(\{1-\frac{y}{1-c}\} + \sqrt{\{1-\frac{y}{1-c}\}^2 + 4 \{\frac{y}{c(1-c)}-1\}}\biggr)/2
\end{equation}
and 
\begin{equation}
\lambda_2 = \biggl(\{1-\frac{y}{1-c}\} - \sqrt{\{1-\frac{y}{1-c}\}^2 + 4 \{\frac{y}{c(1-c)}-1\}}\biggr)/2
\end{equation}
are the two eigenvalues of the transfer matrix $T$. Hence, finally, 
we get [18]
\begin{eqnarray}
\tilde{P}_{2c}(k) = \frac{qy(c-y)}{c(c-qy)} \frac{(\lambda_1^{k-1} - \lambda_2^{k-1})}{(\lambda_1-\lambda_2)} \frac{y(c-qy)}{c(1-c)} \nonumber \\
+ \frac{py}{(c-qy)} \frac{(\lambda_1^{k} - \lambda_2^{k})}{(\lambda_1-\lambda_2)} \frac{y(c-qy)}{c(1-c)} \quad for \quad k \geq 1
\end{eqnarray}
i.e., 
\begin{equation}
\tilde{P}_{2c}(k) = \frac{[py^2c(\lambda_1^k-\lambda_2^k)] + [qy^2(c-y)(\lambda_1^{k-1}-\lambda_2^{k-1})]}{c^2(1-c)(\lambda_1-\lambda_2)} \quad for \quad k \geq 1 
\end{equation}
and 
\begin{equation}
\tilde{P}_{2c}(k) = 1 - \frac{y}{c} \quad for \quad k = 0. 
\end{equation}
The distribution (49)-(50) is in excellent agreement with the corresponding 
results of MC simulation (see figs.5(a) and (b)). 

{\bf 4.2. Distribution of Distance between Jams for $V_{max} > 1$:} 

We have computed the distribution of the distance between jams numerically 
through MC simulation for several values of $V_{max} > 1$; some typical 
distributions for $V_{max} = 5$ are shown in fig.6. The lower is the 
density the more significant is the difference between the distributions 
for $V_{max} = 1$ (fig.5) and for $V_{max} = 5$ (fig.6). This, of course, 
is a consequence of the fact that the higher is the density the fewer 
are the vehicles which can attain speeds larger than unity.

{\bf 5. SUMMARY AND CONCLUSION:} 

In this paper we have calculated the distance headway distribution 
and the distribution of the distance between the jams in the NS model 
of vehicular traffic on single-lane highways. We have derived analytical 
expressions for these distributions in the special case $V_{max}=1$; 
but obtained only numerical results for $V_{max} > 1$ by carrying out 
computer simulation. Our results indicate the coexistence of "free-
flowing" and jammed traffic over a $p$-dependent range of densities 
provided $V_{max} > 1$.  

In the analytical approach followed in this paper one stores the 
information on the state of occupation of each site; therefore, 
this formalism is sometimes called a "site-oriented" approach. 
Very recently a "car-oriented" formalism of analytical calculations 
for the NS model has been developed. Since this new "car-oriented" 
approach takes into account longer range correlations than in the 
"site-oriented" approach, it would be interesting to recalculate 
analytically the distance headway distribution and the distribution 
of distance between the jams in the NS model using this new formalism.

{\bf Acknowledgements:} One of the authors (DC) thanks M. Barma, L. Santen 
and A. Schadschneider for useful discussions, K. Nagel for useful 
correspondence, D. Stauffer for critical comments on the manuscript and  
Alexander von Humboldt Foundation as well as SFB341 K\"oln-Aachen-J\"ulich 
for partial financial supports. DC also thanks D. Stauffer and J. Zittartz 
for a pleasant stay in K\"oln where part of the computational work was 
carried out. 

\newpage

{\bf References} 

[1] D.L. Gerlough and M.J. Huber, {\it Traffic Flow Theory} (National Research Council, Washington, D.C. 1975) 

[2] W. Leutzbach, {\it Introduction to the Theory of Traffic Flow} (Springer, 
Berlin, 1988) 

[3] A.D. May, {\it Traffic Flow Fundamentals} (Prentice-Hall, Englewood Cliffs, 1990) 

[4] I. Prigogine and R. Herman, {\it Kinetic Theory of Vehicular Traffic} (Elsevier, Amsterdam, 1971) 

[5] D.E. Wolf, M. Schreckenberg and A. Bachem (eds.) {\it Traffic and Granular Flow} (World Scientific, Singapore, 1996) 

[6] H. Spohn, {\it Large scale dynamics of interacting particles}
(Springer, Berlin 1991). 

[7] D.C. Gazis, {\it Science}, {\bf 157}, 273 (1967) and references therein.  

[8] K. Nagel, in {\it Physics Computing '92}, R.A. de Groot and 
J. Nadrchal, eds. (World Scientific, Singapore, 1993);\
K. Nagel and M. Schreckenberg, {\it J. Physique I}, {\bf2},
2221 (1992).

[9] S. Wolfram, {\it Theory and Applications of Cellular Automata} (World Scientific, Singapore, 1986) 

[10] M. Schreckenberg, A. Schadschneider, K. Nagel and N. Ito,
{\it Phys. Rev. E}, {\bf 51}, 2939 (1995).

[11] R.B. Stinchcombe, unpublished. 

[12] B. Derrida, M.R. Evans, V. Hakim and V. Pasquier, {\it J. Phys. A}, {\bf 26}, 1493 (1993)  

[13] R.B. Stinchcombe and G.M. Schutz,{\it  Phys. Rev. Lett.} {\bf 75}, 140 (1995)

[14] A. Majumdar, Masters Project (Theory) Report, IIT, Kanpur, 1997 
     (unpublished). 

[15] S. Krauss, P. Wagner and C. Gawron, {\it Phys. Rev. E} {\bf 54}, 3707 
     (1996); see also K. Nagel, {\it Phys. Rev. E} {\bf 53}, 4655 (1996).  

[16] D. Chowdhury, D.E. Wolf and M. Schreckenberg, {\it Physica A}, {\bf 235}, 417 (1997)  

[17] S. L\"ubeck, M. Schreckenberg and K.D. Usadel, Uni-Duisburg Preprint 
     (1997)

[18] S. Sinha, Masters Project (Theory) Report, IIT, Kanpur, 1997, 
     (unpublished). 

[19] A. Schadschneider and M. Schreckenberg, J.Phys.A {\bf 30}, L69 (1997).

\newpage
{\bf Figure Captions:}\\ 

\noindent {\bf Fig.1:} Computer simulation results on the probability 
distribution of the gaps in front of the vehicles in the NS model when 
(a) $V_{max} = 5$ and (b) $V_{max} = 3$;  $p = 0.5$ in both (a) and (b). 
In (a) the symbols $+$, $\Box$, $\diamond$, $\times$ correspond to 
$c = 0.05, 0.08, 0.10$ and $0.20$, respectively. 
In (b) the symbols $+$, $\diamond$, $Box$ and $\times$ correspond to 
$c = 0.10, 0.15, 0.18$ and $0.30$, respectively. 
The lines merely connect the successive data points and serve as a 
guide to the eye. 

\noindent{\bf Fig.2:} Computer simulation results on the probability 
distribution of the gaps in front of the  vehicles in the NS model with 
$V_{max} = 2$ when (a) $p = 0.5$ and (b) $p = 0.9$. 
In (a)  the symbols $\diamond, +, \Box, \times, \triangle$ correspond 
to the densities $c = 0.20, 0.25, 0.30, 0.40$ and $0.60$, respectively. 
In (b)  the symbols $+, \Box, \diamond$ and $\times$ correspond to  
$c = 0.04, 0.06, 0.07$ and $0.09$, respectively. The lines merely 
connect the data points and serve as a guide to the eye. 

\noindent{\bf Fig.3:} Mean-field results on the probability distribution 
of the gaps in front of the vehicles in the NS model with $V_{max} = 1$. 
The symbols $+, \Box, \diamond$ and $\times$ correspond to 
$c = 0.1, 0.2, 0.4$ and $0.6$, respectively.  

\noindent{\bf Fig.4:} Results of 2-cluster theory of the probability 
distribution of the gaps in front of the vehicles in the NS model with 
$V_{max} = 1$. The densities (and symbols) are same as in fig.3 and 
$p = 0.5$.

\noindent{\bf Fig.5(a):} Results of 2-cluster theory of the probability 
distribution of the instantaneous distance between successive jams in 
the NS model with $V_{max} = 1$ when $p = 0.5$. The analytical results 
are plotted as dots connected by lines (solid line, dashed line and 
dashed-dotted line for $c = 0.1, 0.2$ and $0.4$, respectively) whereas 
the results of computer simulation are denoted by the discrete symbols  
($\diamond, +, \Box$ for $c = 0.1, 0.2, 0.4$, respectively). 

\noindent{\bf Fig.5(b):} Same as in fig.5(a), except that the symbols 
$\diamond, +$ and $\Box$ correspond to $c = 0.5, 0.7$ and $0.9$, 
respectively.

\noindent{\bf Fig.6:} Computer simulation results on the probability 
distribution of the instantaneous distance between successive jams in 
the NS model with $V_{max} = 5$ when $p = 0.5$. 
The symbols $\diamond, +$ and $\Box$ correspond to $c = 0.05, 0.10$ and 
$0.90$, respectively.  
The lines merely connect the successive data points and serve as a 
guide to the eye.

\end{document}